\documentclass[preprint]{aastex62}

\usepackage{amsmath}

\usepackage{graphicx}
\received{}
\revised{}
\accepted{}
\shorttitle{Global Solar Magnetic-field and Interplanetary Scintillations}
\shortauthors{Sasikumar Raja et al.}
\begin{document}

\title{Global Solar Magnetic-field and Interplanetary Scintillations During the Past 
Four Solar Cycles}

\correspondingauthor{K.~Sasikumar~Raja}
\email{Sasikumar-Raja.Kantepalli@obspm.fr, sasikumarraja@gmail.com}
\author{K.~Sasikumar~Raja}
\affiliation{Physical Research Laboratory, Navrangpura, Ahmedabad-380 009, India}
\affiliation{LESIA, Observatoire de Paris, Universit\'e PSL, CNRS, Sorbonne Universit\'e, 
Universit\'e de Paris, 5 place Jules Janssen, 92195 Meudon, France}
\author{P.~Janardhan}
\affiliation{Physical Research Laboratory, Navrangpura, Ahmedabad-380 009, India}
\author{Susanta~Kumar~Bisoi}
\affiliation{Key Laboratory of Solar Activity, National Astronomical Observatories, Chinese Academy of Sciences, Beijing 100 012, People's Republic of China}

\author{Madhusudan~Ingale}
\affil{Indian Institute of Science Education and Research, Pashan, Pune - 411 008, India}

\author{Prasad~Subramanian}
\affil{Indian Institute of Science Education and Research, Pashan, Pune - 411 008, India}

\author{K.~Fujiki}
\affil{Institute for Space-Earth Environmental Research, Nagoya, Japan}

\author{Milan~Maksimovic}
\affiliation{LESIA, Observatoire de Paris, Universit\'e PSL, CNRS, Sorbonne Universit\'e, 
Universit\'e de Paris, 5 place Jules Janssen, 92195 Meudon, France}




\begin{abstract}

The extended minimum of Solar Cycle 23, the extremely quiet solar-wind conditions prevailing and the mini-maximum of Solar Cycle 24 drew global attention and many authors have since attempted to predict the amplitude of the upcoming Solar Cycle 25, which is predicted to be the third successive weak cycle; it is a unique opportunity to probe the Sun during such quiet periods. 
Earlier work has established a steady decline, over two decades, in solar photospheric fields at latitudes above $45^{\circ}$ and a similar decline in solar-wind micro-turbulence levels as measured by interplanetary scintillation (IPS) observations. However, the relation between the photospheric magnetic-fields and those in the low corona/solar-wind are not straightforward. Therefore, in the present article, we have used potential-field source-surface (PFSS) extrapolations to deduce global magnetic-fields using synoptic magnetograms observed with National Solar Observatory (NSO), Kitt Peak, 
USA (NSO/KP) and \emph{Solar Optical Long-term Investigation of the Sun} (NSO/SOLIS) instruments  during 1975\,-- \,2018. Furthermore, we have measured the normalized scintillation index [m] using the IPS observations carried out at the Institute of Space Earth Environment Research (ISEE), Japan during 1983\,-- \,2017. From these observations, we 
have found that, since the mid 1990s, the magnetic-field over different latitudes at 2.5 $\rm R_{\odot}$ and 10 $\rm R_{\odot}$(extrapolated using PFSS method) has
decreased by $\approx 11.3-22.2\,\%$. In phase with the declining magnetic-fields, the quantity m also declined by $\approx 23.6\,\%$. These observations emphasize the inter-relationship among the global magnetic-field and various turbulence parameters
in the solar corona and solar-wind.\\

\end{abstract}


\section{Introduction}\label{S-Introduction} 

The magnetic-field plays a crucial role in understanding the various 
phenomenon that occur on the Sun and solar atmosphere. 
Most of the features (\emph{e.g.} sunspots, filaments, prominences, coronal holes)
and transient events (\emph{e.g.} solar flares, coronal mass ejections, and non-thermal 
radio bursts) are related to magnetic-field. Recent work has established a 
steady decline, over the past two decades, in solar photospheric fields at latitudes 
above $45^{\circ}$ and a similar decline in solar-wind micro-turbulence
levels as measured by interplanetary scintillation (IPS) 
observations \citep{Jan2010, Jan2011, Jan2015}. 
Solar cycle 24 has also shown a significant 
asymmetry in the times of reversals of the polar field in the two solar hemispheres 
\citep{Jan2018}, leading to speculation as to whether we are heading towards a another 
Grand solar minimum like the Maunder minimum if the decline in the photospheric fields 
will continue beyond 2020, the expected minimum of the current Solar Cycle 24.

Although photospheric magnetic-fields is available from many decades \citep[see for example][]{Hale1908}, our understanding of the coronal magnetic-fields is limited \citep{Lin2000}.
Direct methods such as the Zeeman effect \citep{Harvey1969} and 
Hanle effect \citep{Mickey1973,Querfeld1984, Arnaud1987} fail in the low density corona. 
Magnetic-field measurements in the outer corona derived 
using Faraday rotation observations are limited 
due to various (observational and instrumental) constraints \citep{Stelzried1970,Bird1981,Bird1982}.
Magnetic-field measurements that are reported using a few indirect techniques using 
polarization observations of solar radio bursts \citep{Sastry2009, 
Ramesh2010b, Sas2013, Sas2014} are rare. 
Therefore, in the inner corona, magnetic-field measurements are limited to 
extrapolation techniques \citep{Schatten1969,Schrijver2003}. 

In this article, we use the photospheric synoptic magnetogram data observed 
using National Solar Observatory (NSO), Kitt Peak, 
USA (NSO/KP) and \emph{Solar Optical Long-term Investigation of the Sun} (NSO/SOLIS)
instruments. We used potential-field source surface
(PFSS) extrapolation routines available in the IDL/solarsoft library \citep{Free1998} to extrapolate the magnetic-fields to 2.5 and 10 $\rm R_{\odot}$\citep{Altschuler1969,Sca1969,Hoek1984,Wan1992,Sch2003}
to examine if the extended decline in the photospheric fields (maybe high latitude photospheric fields?) is mirrored in the coronal fields as well.
%to see if the extended decline 
%in solar photospheric fields can be seen using just PFSS extrapolations. 
The magnetogram data that we used was observed 
during 1975\,--\,April 2018. On the other hand, we used the observations of inter-planetary scintillation (IPS) carried out at Institute for Space-Earth Environmental Research (ISEE), Japan during 1983\,--\,2017. 
Using the IPS observations of 27 radio sources carried out in the 
heliocentric distance 0.2\,--\,0.8 AU (astronomical unit, 1 AU = 215 $\rm R_{\odot}$), we measured the normalized scintillation index [m]. 
In this article, we examine the relation between the global magnetic-fields (at the photospheric level and in the inner solar-wind) and the level of interplanetary scintillations (characterized by m).
%In the article, we reported the way the global magnetic-field (at photosphere and 
%in solar-wind) and the quantity `m' (in the solar-wind) are inter-related. 

The observational details of magnetograms and interplanetary scintillations are 
discussed in Section \ref{sec:observations}. The PFSS extrapolation technique, measurements 
of magnetic-fields over different latitudes, and the normalized scintillation index [m] are discussed in Section \ref{sec:data_analysis}. 
The observational results and discussions are described in Section \ref{sec:results}. 
The summary and conclusions are given in Section \ref{sec:conclusions}.

\section{Observations}\label{sec:observations}

The observational details of magnetogram data, IPS observations, and sunspot number are 
discussed in this section. 

\subsection{Magnetogram Data}\label{sec:obsB}
In the present study, we use the synoptic magnetograms from NSO/KP 
observed during 1975.13\,--\,2003.66.
This duration corresponds to the Carrington Rotations (CR) CR1625 to 
CR2006. Since 2003.66, such observations are carried out using  
Vector Stokes Magnetograph, one of the three instruments that comprises SOLIS. During CR2007 and CR2206, we used the data observed using the SOLIS instrument.
The synoptic maps are prepared using the full-disk magnetograms (see Figure \ref{fig:hairy_ball})
observed over a Carrington rotation. The synoptic maps 
were stored in the \textsf{FITS} format. The data are stored as a
$180\times360$ array format. This means that the resolution of a
synoptic map is $1^\circ$ in both longitudinal ($0^{\circ}-360^\circ$) and 
latitudinal ($-90^\circ$ to $90^{\circ}$) directions. 
The full-disk magnetograms are mapped into longitude and latitude coordinates 
and added together to form the final synoptic magnetogram (see upper panel of the 
Figure \ref{fig:mg}).

\subsection{Interplanetary Scintillation data}\label{sec:obsM}
Inter-planetary scintillation (IPS) or intensity scintillation observation is a
well-established technique to probe the solar-wind in the
inner heliosphere. IPS is basically a diffraction phenomenon in which coherent 
electromagnetic radiation
from a distance radio source experiences the scattering when it is observed through the turbulent and 
refracting solar-wind and thus the temporal variation of the flux density 
when observed from Earth \citep{Hew1964, Ana1980, Koj1990, Jan1993, Jan1996, Asa1998,
Man2010, Tok2010}. 

ISEE has been carrying out IPS observations using a three station facility 
located at Fuji (long. $138^{\circ} 36'42''$ E and lat. $35^{\circ} 25'36''$ N), Toyokawa (long. $137^{\circ}22'09''$ E	and lat. $34^{\circ}50'05''$ N), Sugadaira (long. $138^{\circ}19'16''$ E and lat. $36^{\circ}31'12''$ N) during 1983-1994. In addition, a fourth station has been commissioned at Kiso (long. $137^{\circ}37'49''$ E and lat. $35^{\circ}47'34''$ N) in the year 1994 and then onwards a four station facility has been used to measure the 
solar-wind speed by using cross-correlation analysis. The current four station 
network provide the more robust estimates of the solar-wind speed owing to the redundancy in the baseline geometry. 
However, the scintillation index was measured using the telescope located at Fuji during 1983-1994. After 1994, a new facility located at Kiso has been used to measure the scintillation index. 
We note here that the telescopes located at all four stations are identical.

\subsection{Sunspot Number}

In this work, we make use of the revised sunspot numbers 
(ver-2.0)(\url{www.sidc.be/silso/newdataset}) prepared by re-calibrating the sunspots that were observed over 400 years \citep{Cle2015}. 

\section{Data Analysis}\label{sec:data_analysis}

In this section we describe the PFSS extrapolation technique and the way that we measured the averaged magnetic-fields 
over different latitude regions using the synoptic magnetogram data (see Section \ref{sec:pfss_123} and Section \ref{sec:mag_123}). 
In addition, we discuss the way that we measured the normalized scintillation index 
using the IPS observations (see Section \ref{sec:ips_123}).

\subsection{PFSS Extrapolation}\label{sec:pfss_123}

\begin{center}
\begin{figure*}
\begin{tabular}{cccc}
\includegraphics[width=0.45\textwidth]{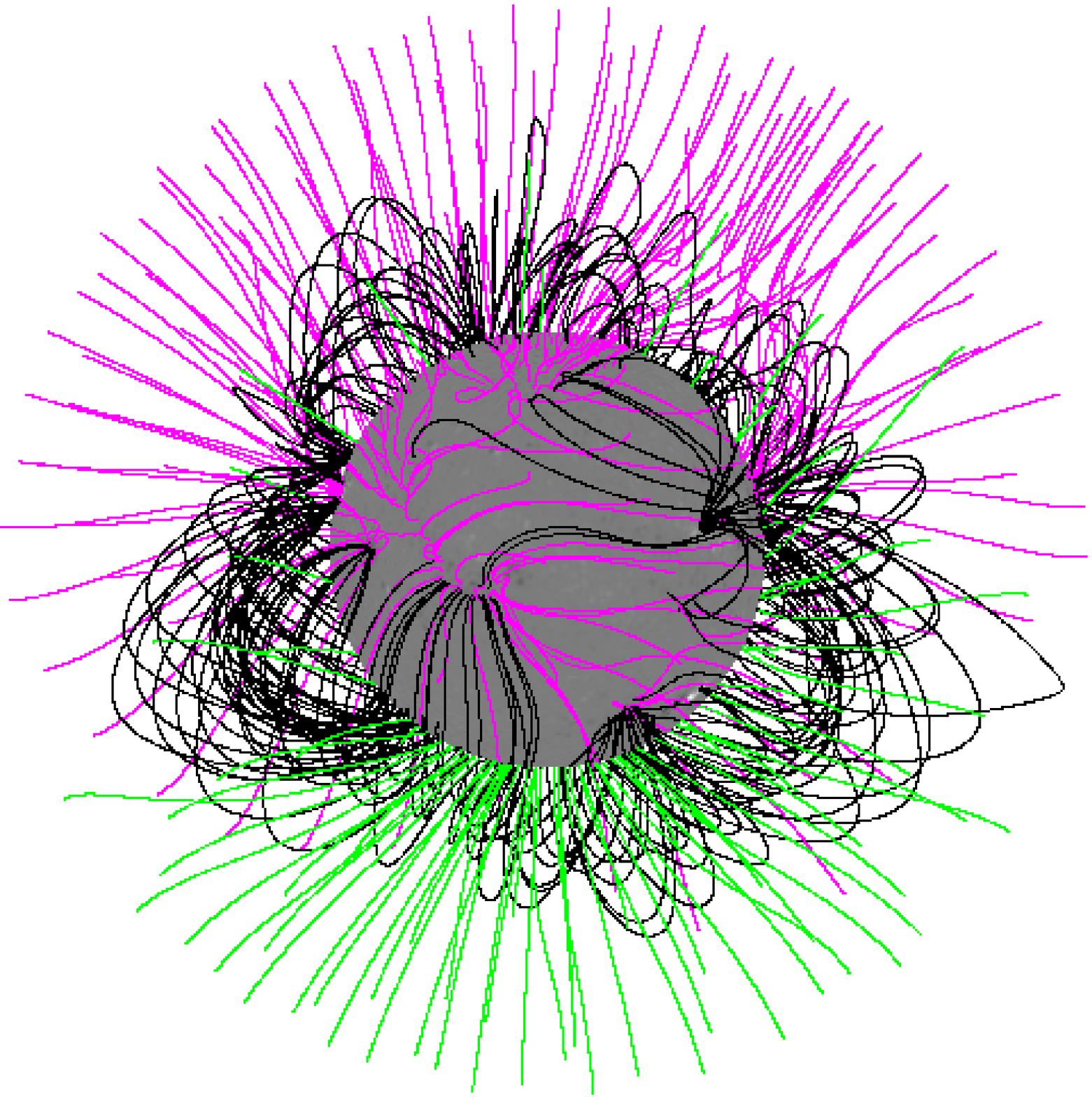} &
\includegraphics[width=0.45\textwidth]{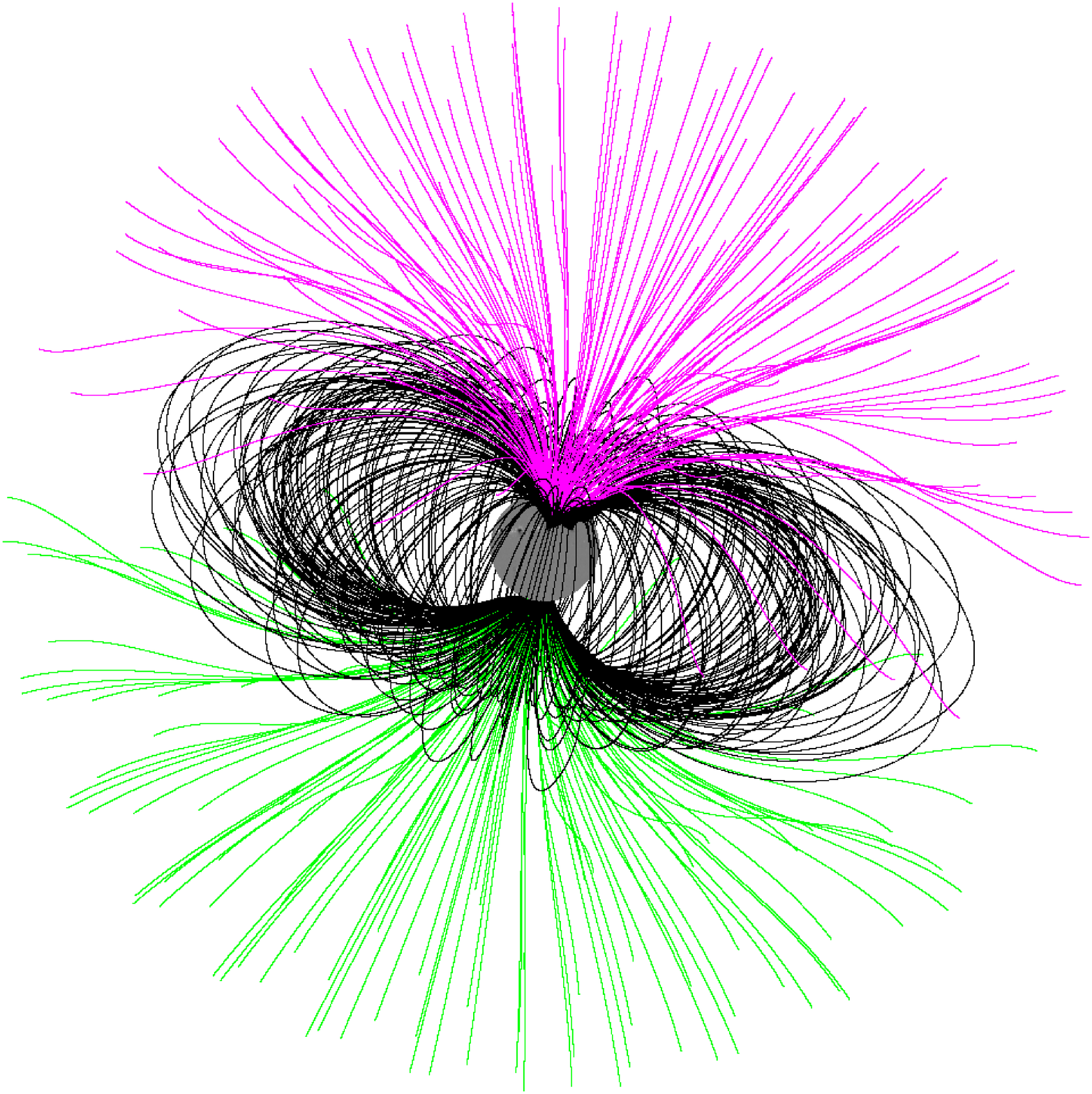} \\
 (a) & (b)\\
\includegraphics[width=0.45\textwidth]{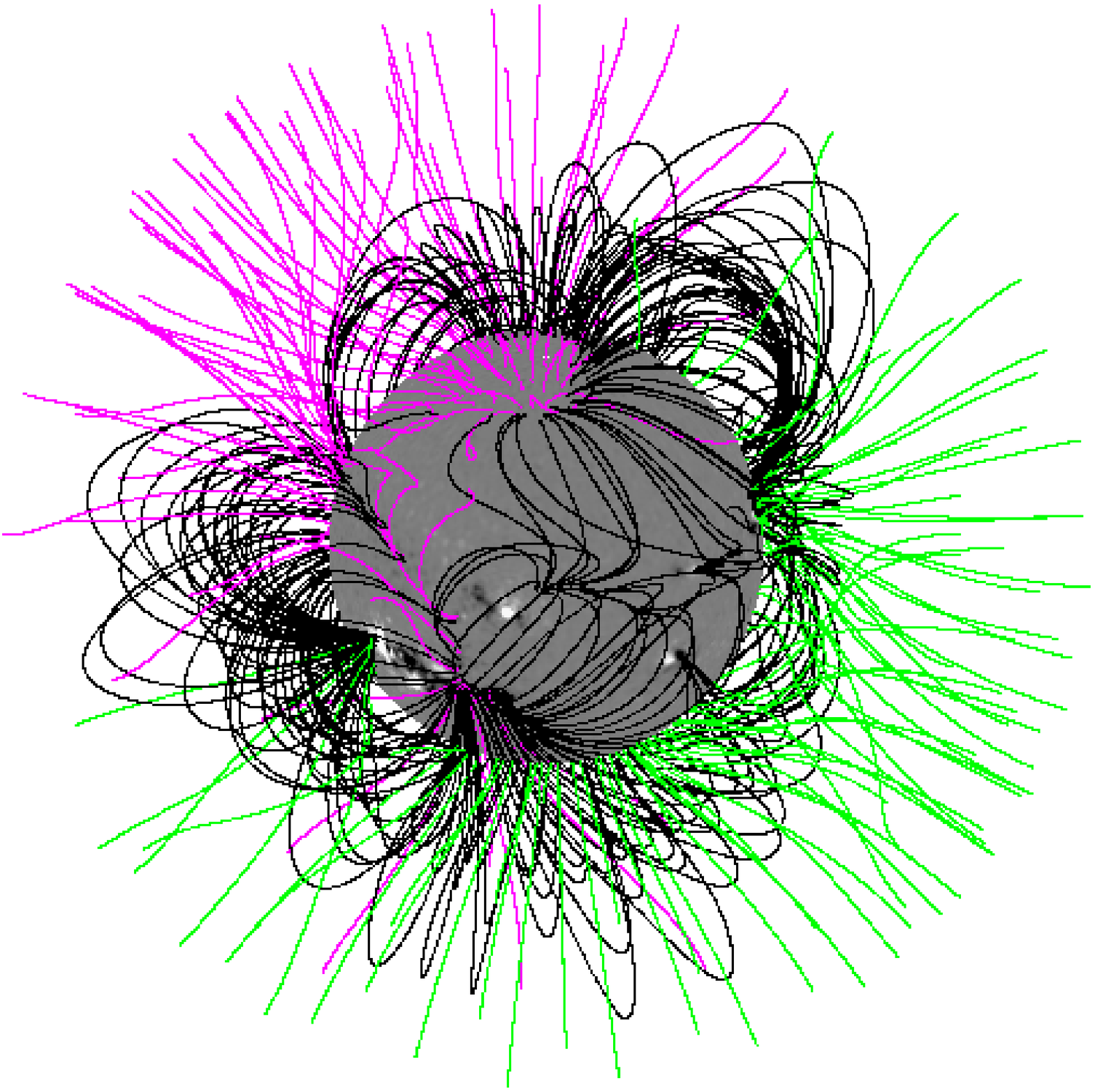} &
\includegraphics[width=0.45\textwidth]{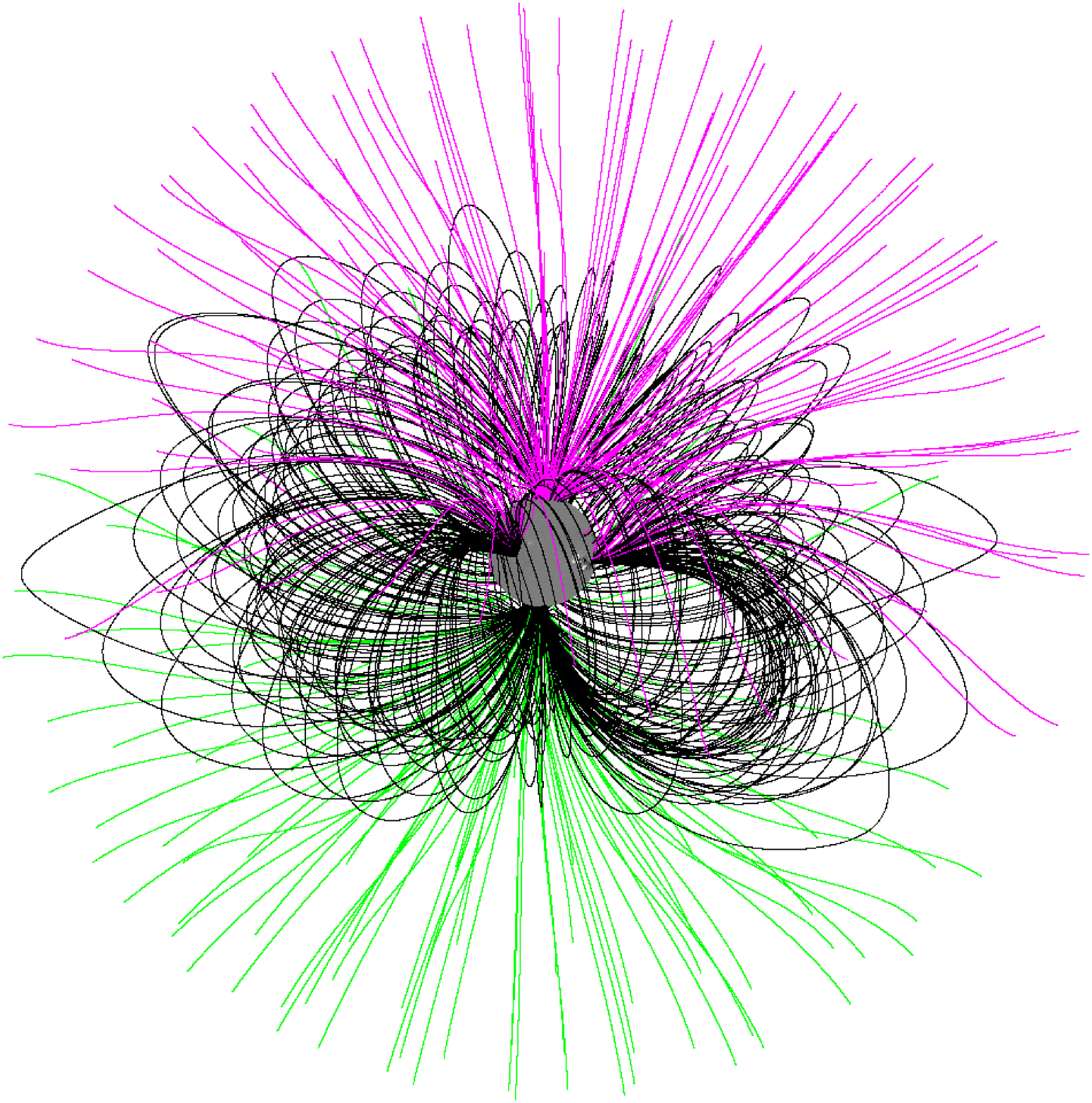} \\
 (c) & (d)\\
 \end{tabular}
\caption{The {\bf upper (a and b)} and {\bf lower (c and d) panels} represents the PFSS extrapolated magnetic-fields derived from the full disk magnetograms observed on 2011 August 9 (solar minimum) and 2004 July 18 (solar maximum) respectively. 
The {\bf left (a and c)} and {\bf right (b and d)} panels show the extrapolated magnetic-fields 
from 1.5\,--\,2.5 $\rm R_{\odot}$ and 5\,--\,10  $\rm R_{\odot}$ respectively. 
The \emph{gray colored disk} at the center is the magnetogram observed at NSO/KP or NSO/SOLIS instruments. The \emph{black lines} represent the closed field lines. The lines, respectively, in \emph{majenta and green} indicate the negative and positive polarities of the open magnetic-field lines.}
\label{fig:hairy_ball}
\end{figure*}
\end{center}

The photospheric magnetic-field and its spatial distribution are
routinely observed using the magnetographs. However 
coronal magnetic-fields are challenging to probe owing to the 
low coronal density, as previously mentioned. Therefore, global magnetic-fields in the corona are commonly 
modeled using the potential-field source-surface (PFSS) model \citep{Schatten1969}.
The upper (a and b) and lower (c and d) panels of Figure \ref{fig:hairy_ball} show the so called ``hairy Sun'' images observed on 2011 August 9 (during solar minimum) and 18 July 2004 (during solar maximum) respectively. Similarly the left (a and c) and right (b and d) panels represents  
the extrapolated field lines drawn from 1.5\,--\,2.5 $\rm R_{\odot}$ and 5\,--\,10 $\rm R_{\odot}$ respectively using the PFSS model. The field lines in black
are closed, \emph{i.e.} they intersect the inner boundary (\emph{i.e.} the photosphere)
in two places. The field lines in magenta and green colors are open, 
\emph{i.e.} they intersect both inner and outer boundaries (the source surface) 
of the model. The magenta and green colors indicate the 
negative or positive polarities 
respectively (\url{www.lmsal.com/~derosa/pfsspack/}).

\begin{figure}
   \centerline{\includegraphics[width=1.0\textwidth,clip=]{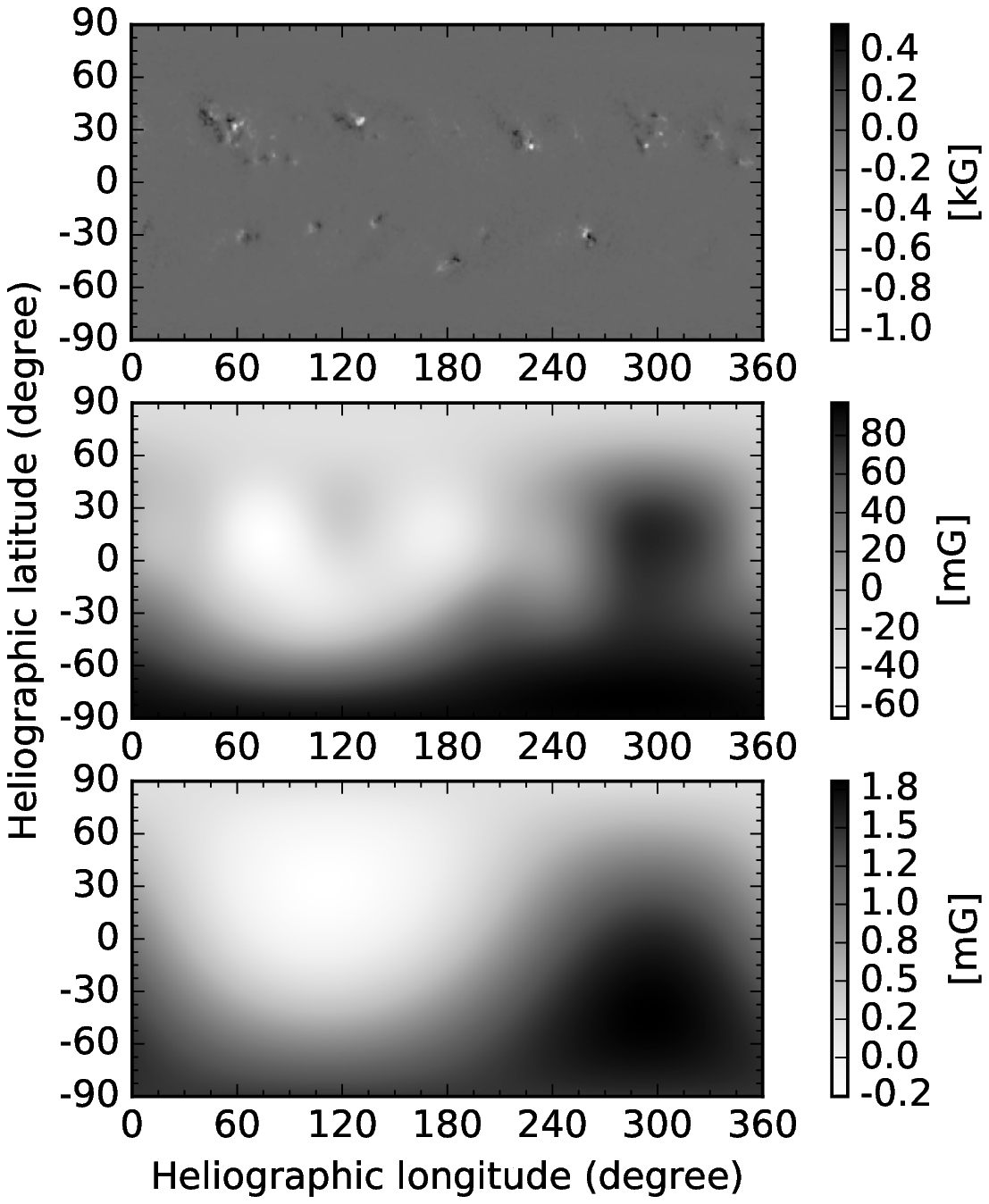}
              }
              \caption{The synoptic magnetogram ({\bf upper panel}) was observed during the CR2114 using NSO/SOLIS instrument at the wavelength 630.150 nm. The {\bf upper panel} shows the distribution of observed photospheric 
              magnetic-fields. The {\bf middle} and {\bf lower panels}
              are the extrapolated synoptic magnetograms (of the one in the upper panel)
              to the source surface 2.5 and 10 $R_\odot$ respectively.}
\label{fig:mg}
\end{figure}

In the present work, we used the synoptic magnetograms observed at NSO/KP and NSO/SOLIS 
instruments and extrapolated the magnetic-field to 2.5 and 10 $\rm R_{\odot}$
using the PFSS model. The key assumption of this model is that there is 
zero electric current in the solar corona. Usually this method is applied up to the 
heliocentric distance $2.5~\rm R_{\odot}$. Beyond this distance, in general, the magnetic-fields are
radial and therefore, we extrapolated further to $10~\rm R_{\odot}$.
The upper panel of Figure \ref{fig:mg} shows the observationally derived synoptic magnetogram at photospheric height. The middle and lower panels are the 
extrapolated magnetograms to the heliocentric distances 2.5 and 10 $\rm R_{\odot}$.  
Note that we used this model because it is one of the basic and routinely used 
models when compared to other models such as the current-sheet source-surface (CSSS) model \citep{Zhao1995} and non-linear force-free models \citep{Ball2000, Mac2006}.

\subsection{Magnetic-field Measurements}\label{sec:mag_123}
\begin{figure}[!ht]
                                % includes the two top panels 
   \centerline{\hspace*{0.015\textwidth}
               \includegraphics[width=0.55\textwidth,clip=]{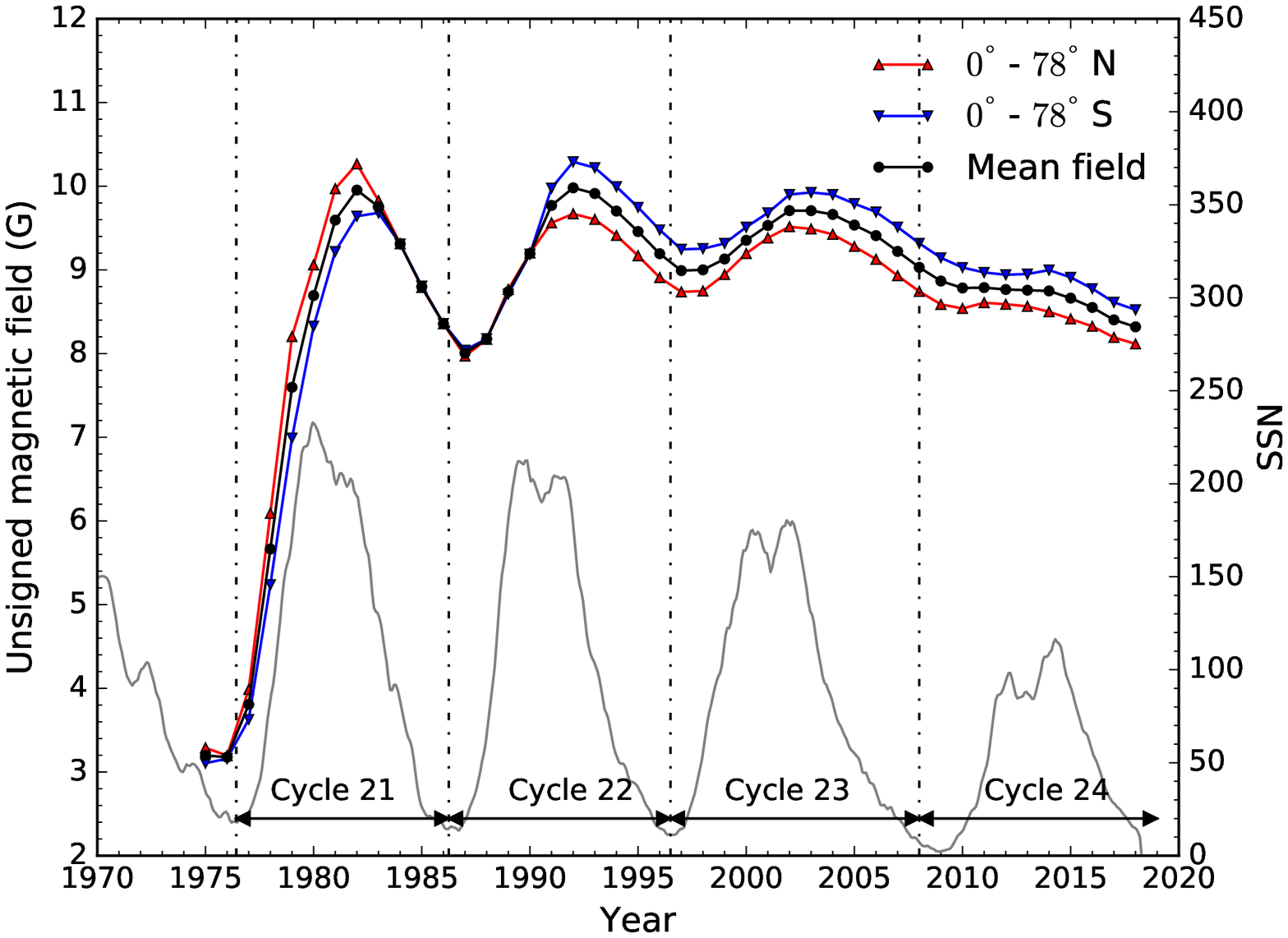}
               \hspace*{-0.03\textwidth}
               \includegraphics[width=0.55\textwidth,clip=]{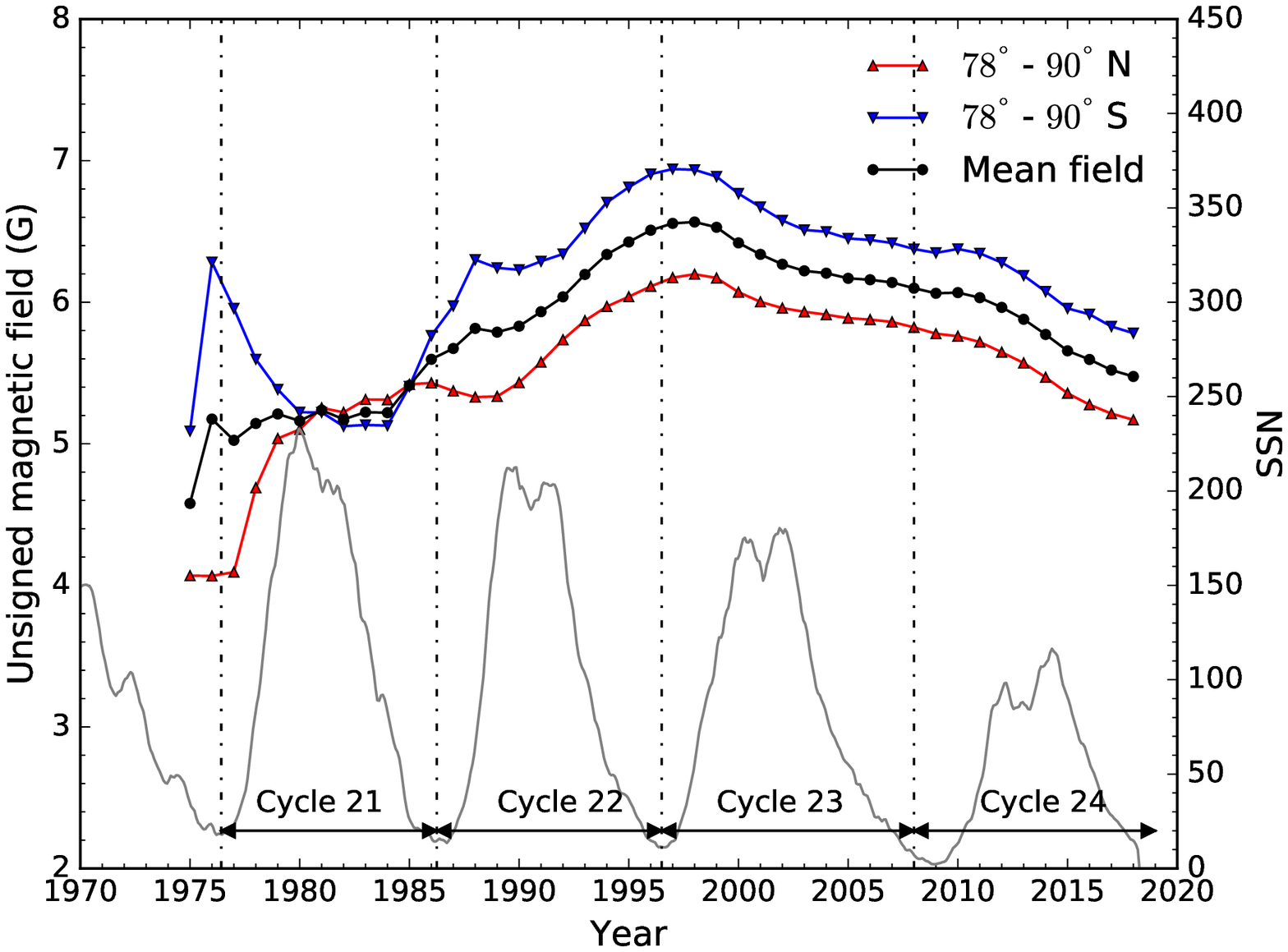}
              }
     \vspace{-0.375\textwidth}   % Shift close to the panel top 
     \centerline{\Large \bf     % Includes the labels (here needs the color 
                                %   package, see beginning of this file)
      \hspace{0.025 \textwidth}  \color{black}{(a)}
      \hspace{0.46\textwidth}  \color{black}{(b)}
         \hfill}
     \vspace{0.33\textwidth}    % Shift back to the panel bottom 
   \centerline{\hspace*{0.015\textwidth}
               \includegraphics[width=0.55\textwidth,clip=]{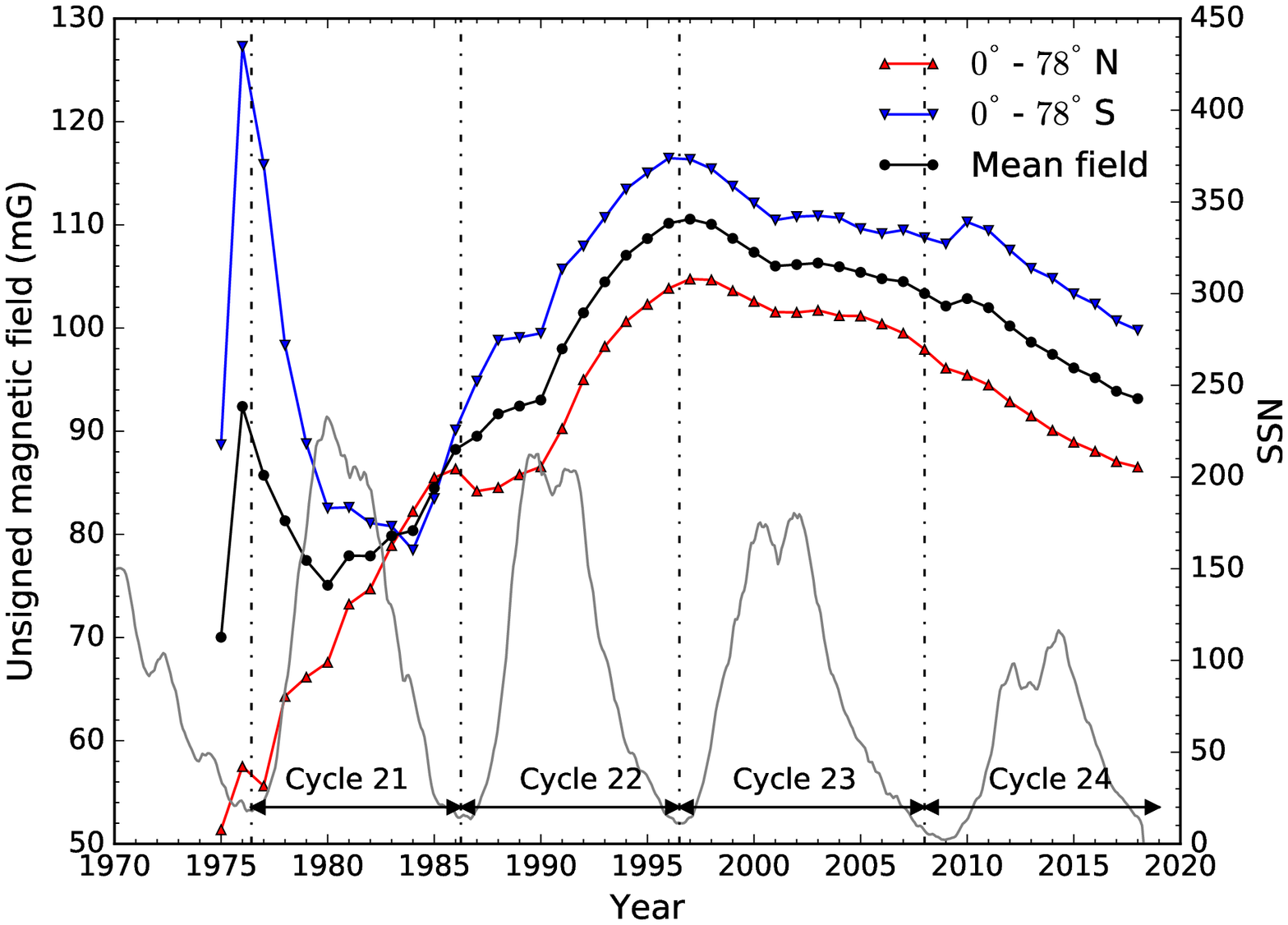}
               \hspace*{-0.03\textwidth}
               \includegraphics[width=0.55\textwidth,clip=]{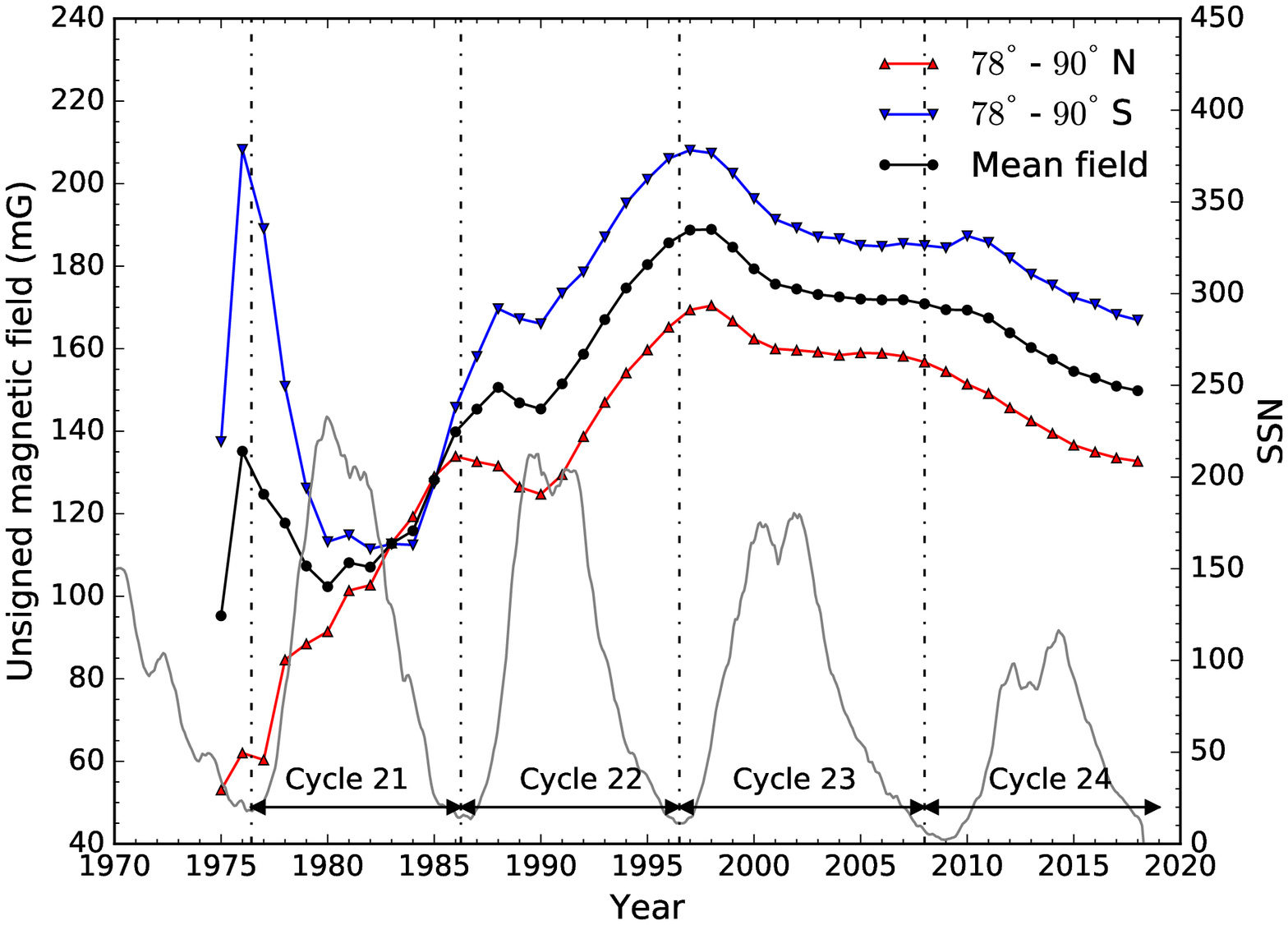}
              }
     \vspace{-0.375\textwidth}   % Shift close to the panel top 
     \centerline{\Large \bf     % Includes the labels (here needs the color package)
      \hspace{0.025 \textwidth} \color{black}{(c)}
      \hspace{0.46\textwidth}  \color{black}{(d)}
         \hfill}
     \vspace{0.33\textwidth}    % Shift back to the panel bottom 
              
      \centerline{\hspace*{0.015\textwidth}
               \includegraphics[width=0.56\textwidth,clip=]{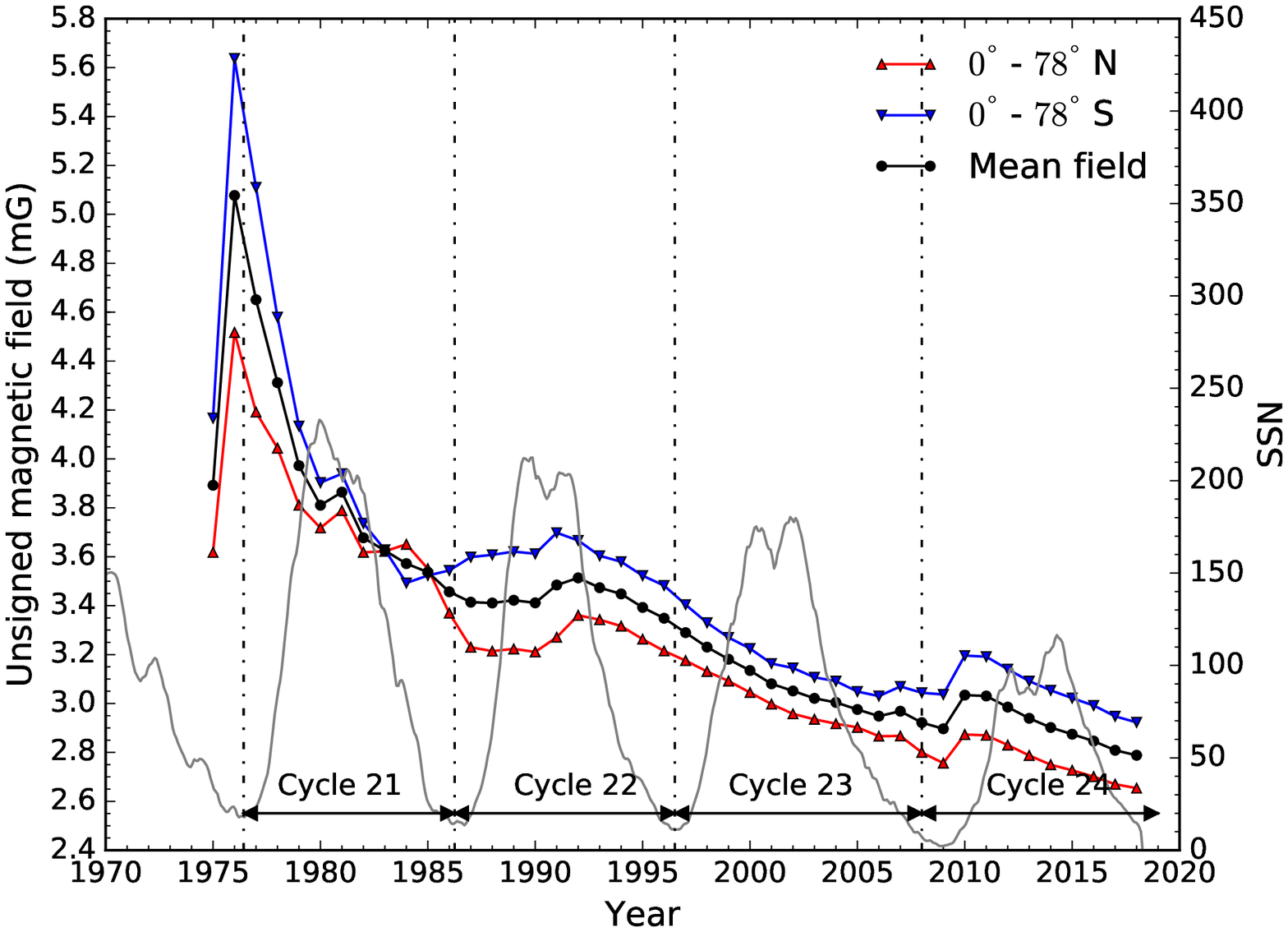}
               \hspace*{-0.03\textwidth}
               \includegraphics[width=0.56\textwidth,clip=]{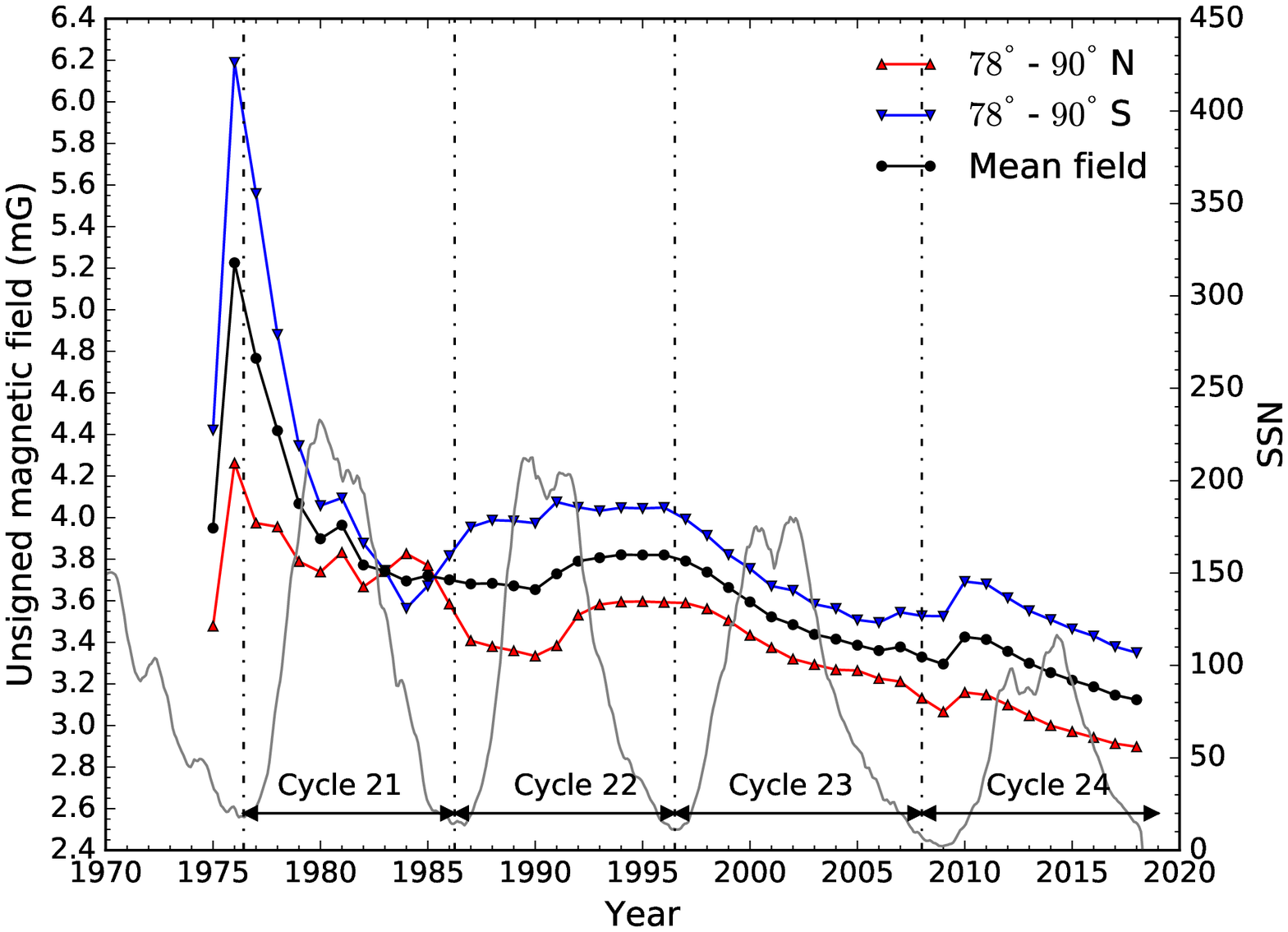}
              }
     \vspace{-0.375\textwidth}   % Shift close to the panel top 
     \centerline{\Large \bf     % Includes the labels (here needs the color package)
      \hspace{0.02 \textwidth} \color{black}{(e)}
      \hspace{0.47\textwidth}  \color{black}{(f)}
         \hfill}
     \vspace{0.32\textwidth}    % Shift back to the panel bottom 

\caption{Variation of magnetic-field with the solar cycle. 
The {\bf left \bf (a, c, e)} and {\bf right} columns {\bf (b, d, f)} show the average of Region-A and poloidal fields respectively. 
In each panel the \emph{triangles pointing upward and triangles pointing downward} indicate the 
northern and  southern hemispheric fields.  The \emph{circle in black} indicate the average of both northern and southern hemispheric fields. The \emph{gray solid line} shows the monthly averaged sunspot number (SSN). Panels {\bf a} and {\bf b} shows the field on photosphere.
The panels {\bf c} and {\bf d} show the extrapolated field at $2.5 R_{\odot}$. 
Panels {\bf e} and {\bf f} show the extrapolated field at 10 $\rm R_{\odot}$.}
\label{fig:extrap1}
\end{figure}

\begin{figure}[!ht]
                                % includes the two top panels 
   \centerline{\hspace*{0.015\textwidth}
               \includegraphics[width=0.57\textwidth,clip=]{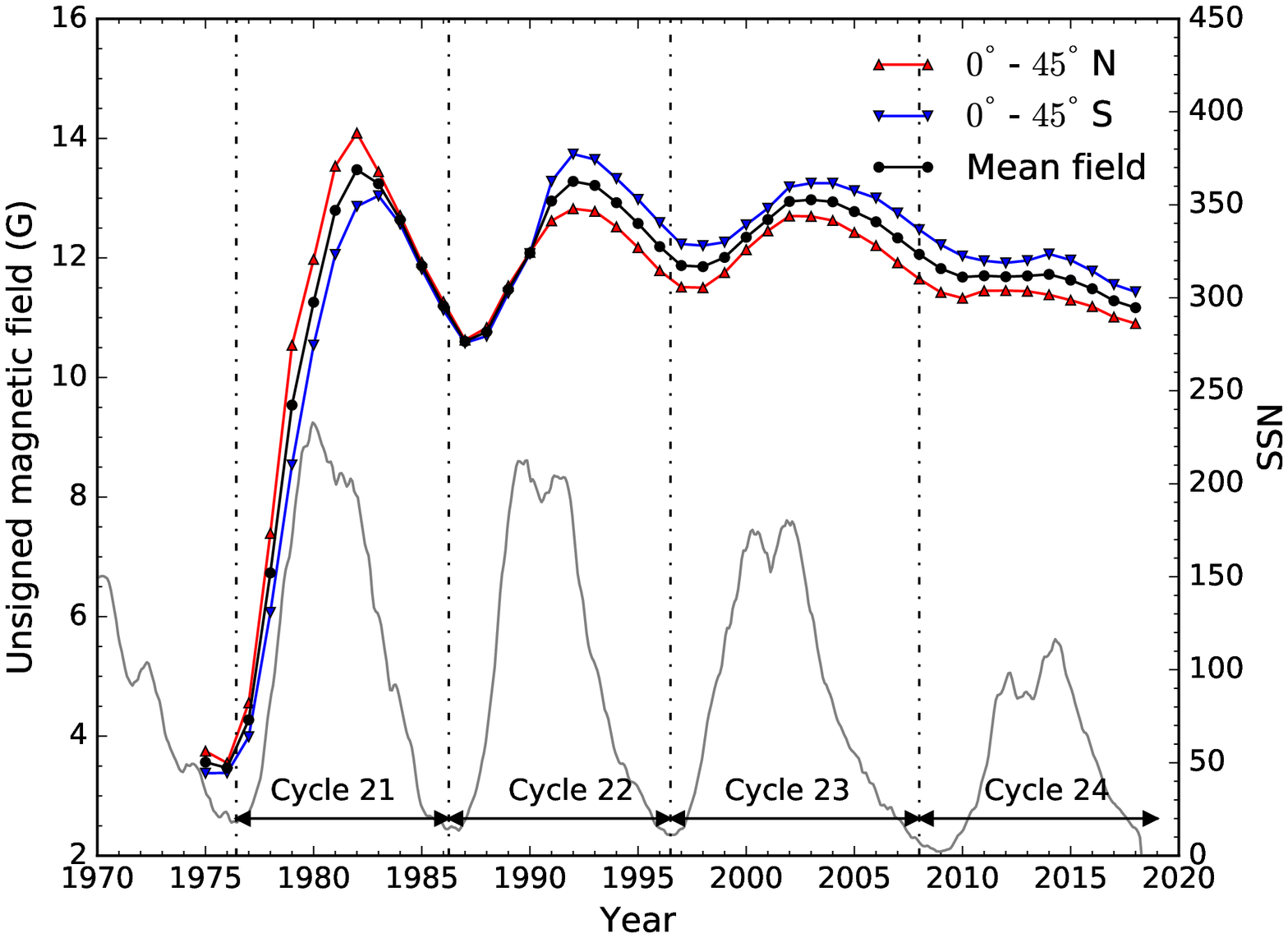}
               \hspace*{-0.03\textwidth}
               \includegraphics[width=0.57\textwidth,clip=]{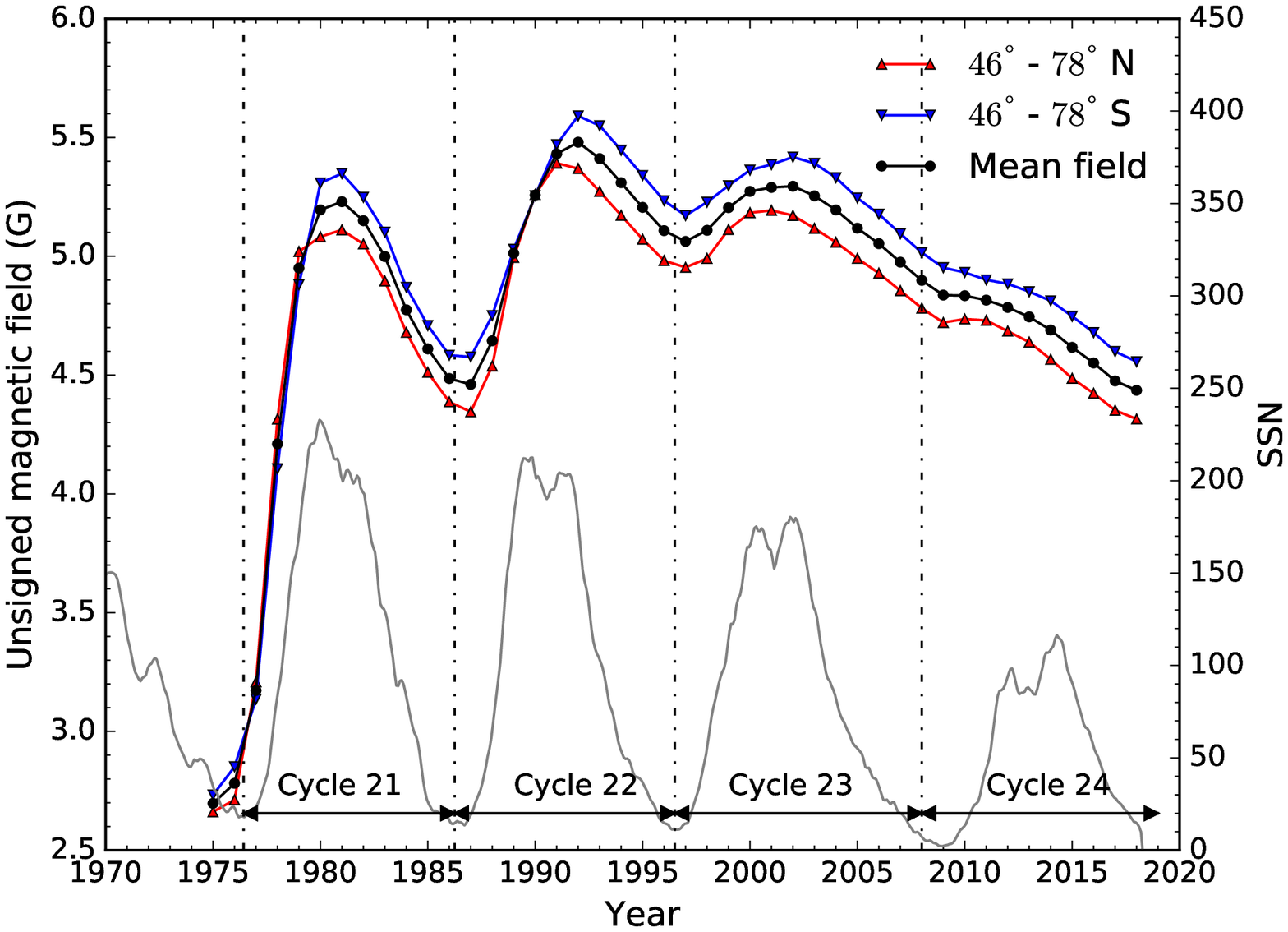}
              }
     \vspace{-0.375\textwidth}   % Shift close to the panel top 
     \centerline{\Large \bf     % Includes the labels (here needs the color 
                                %   package, see beginning of this file)
      \hspace{0.020 \textwidth}  \color{black}{(a)}
      \hspace{0.47\textwidth}  \color{black}{(b)}
         \hfill}
     \vspace{0.33\textwidth}    % Shift back to the panel bottom 
   \centerline{\hspace*{0.015\textwidth}
               \includegraphics[width=0.57\textwidth,clip=]{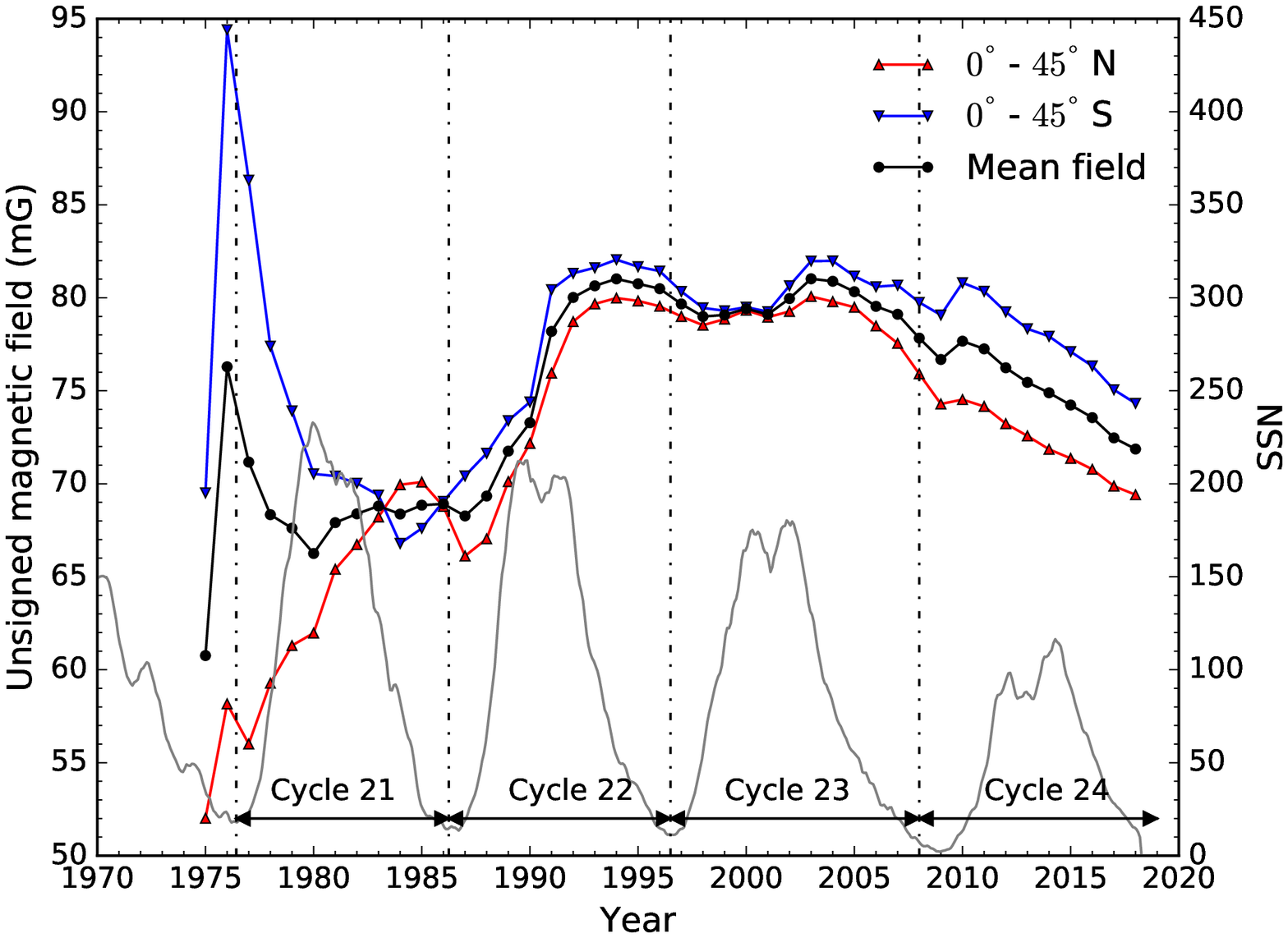}
               \hspace*{-0.03\textwidth}
               \includegraphics[width=0.57\textwidth,clip=]{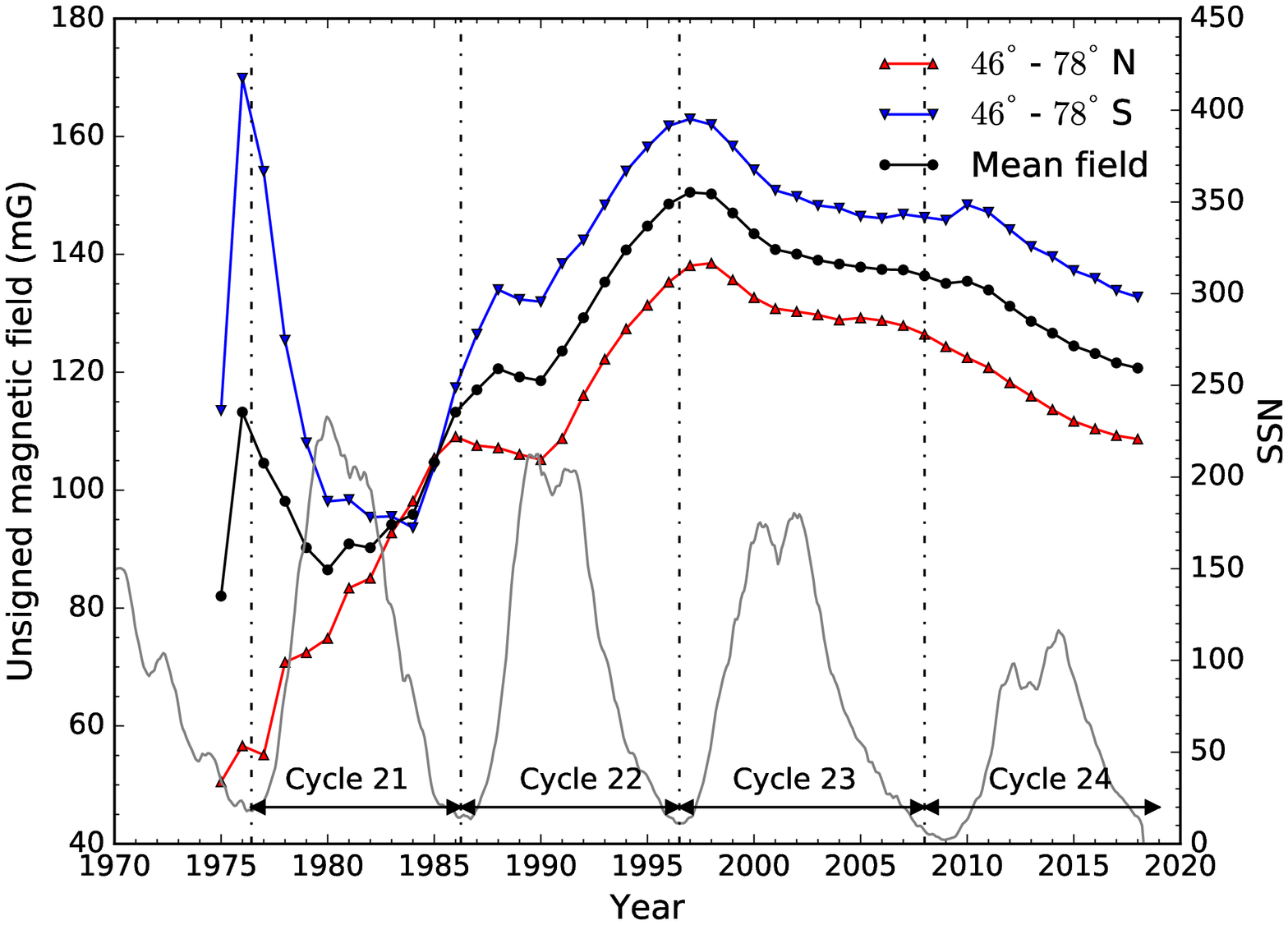}
              }
     \vspace{-0.375\textwidth}   % Shift close to the panel top 
     \centerline{\Large \bf     % Includes the labels (here needs the color package)
      \hspace{0.02\textwidth} \color{black}{(c)}
      \hspace{0.47\textwidth}  \color{black}{(d)}
         \hfill}
     \vspace{0.33\textwidth}    % Shift back to the panel bottom 
              
      \centerline{\hspace*{0.015\textwidth}
               \includegraphics[width=0.575\textwidth,clip=]{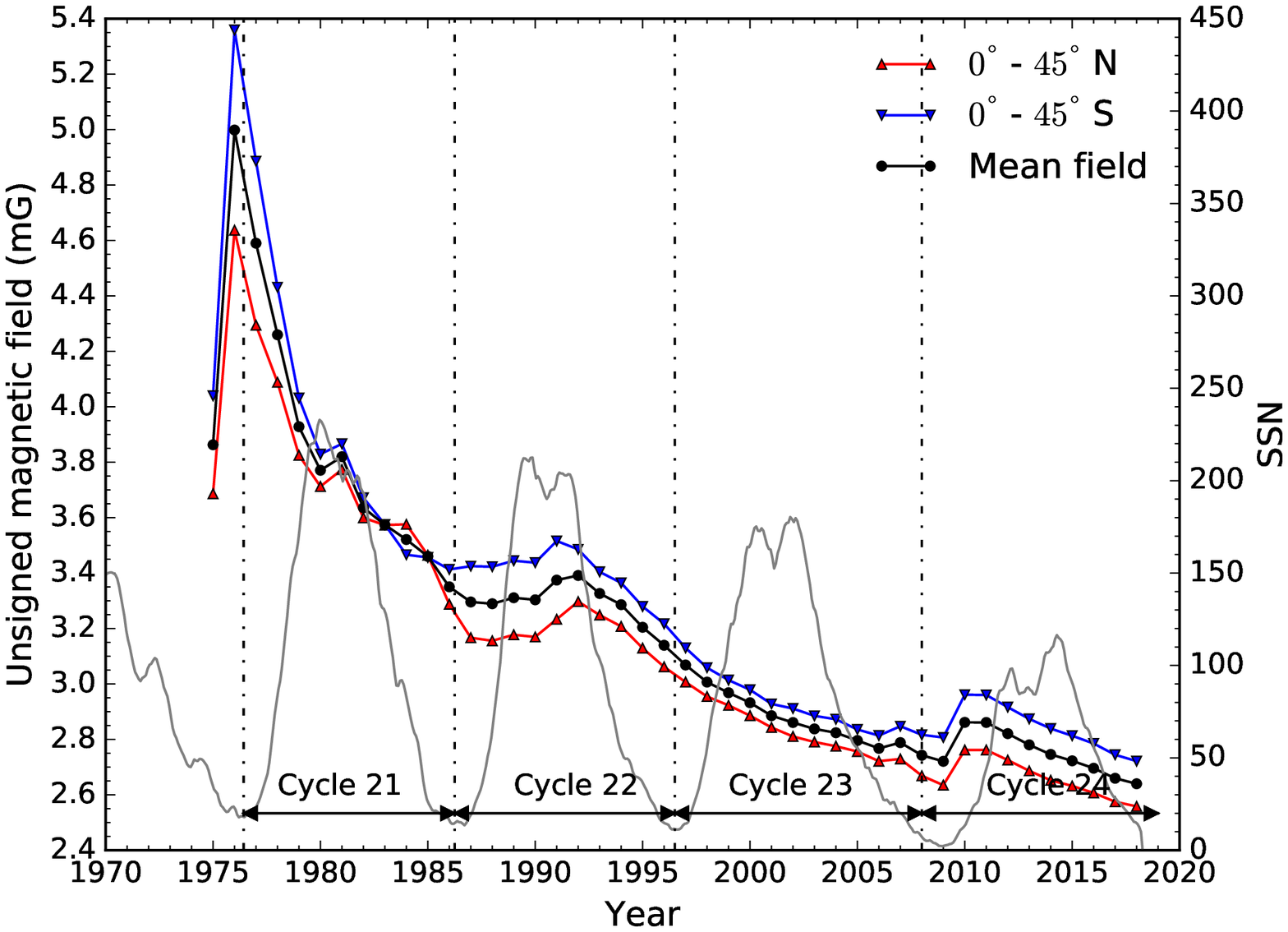}
               \hspace*{-0.03\textwidth}
               \includegraphics[width=0.575\textwidth,clip=]{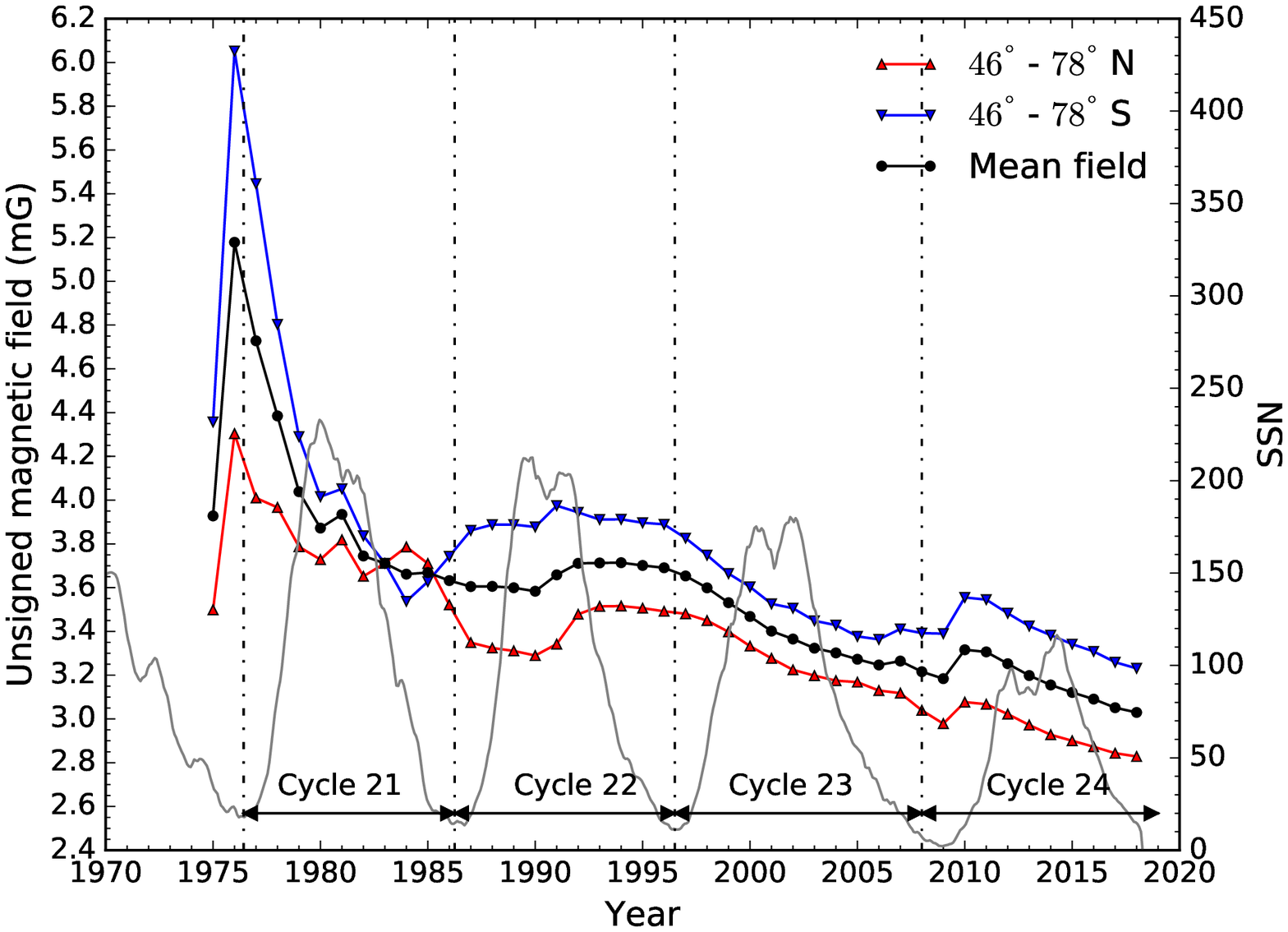}
              }
     \vspace{-0.375\textwidth}   % Shift close to the panel top 
     \centerline{\Large \bf     % Includes the labels (here needs the color package)
      \hspace{0.02 \textwidth} \color{black}{(e)}
      \hspace{0.47\textwidth}  \color{black}{(f)}
         \hfill}
     \vspace{0.32\textwidth}    % Shift back to the panel bottom 

\caption{Variation of magnetic-field with the solar cycle. 
The {\bf left} {\bf (a, c, e)} and {\bf right} columns {\bf (b, d, f)} show the toroidal and mid-latitude fields respectively. In each panel the \emph{triangles pointing upward and triangles pointing downward} indicate the northern and  southern hemispheric fields.  The \emph{circle in black} indicate the average of both northern and southern hemispheric fields. The \emph{gray solid line} shows the monthly averaged sunspot number (SSN). Panels (a) and (b) shows the field on photosphere. The panels {\bf c} and {\bf d} show the extrapolated field at $2.5 R_{\odot}$. Panels {\bf e} and {\bf f} show the extrapolated field at 10 $\rm R_{\odot}$.}
\label{fig:extrap2}
\end{figure}

Using the magnetograms observed at photospheric height and 
the extrapolated magnetograms at $2.5$ and $10~R_{\odot}$, we have studied the 
magnetic-field variations over different range of latitudes.
As we are interested in latitudinal variation of the magnetic-field, 
we find an averaged magnetic-field along longitudes [$\phi_{i,n}$] using

\begin{equation}\label{eq:1d}
 \phi_{i,n}={\Sigma_{j=1}^{360} ~ \phi_{i,j,n} \over 360}
\end{equation}

\noindent where i, j and n are the latitude, longitude, and CR number. 
After the average, the size of the array reduces to $180 \times 1$. 
Then we measured the averaged magnetic-field [$\phi_n$] over 
selected latitude intervals for a given CR number [n] using

\begin{equation}\label{eq:2d}
 \phi_n={{\Sigma_{i=k}^p \phi_{i,n}} \over {p-k+1}}
\end{equation}

\noindent where k and p are the row numbers corresponding to the selected latitude bin.
Using the Equations \ref{eq:1d} and \ref{eq:2d}, we have measured the magnetic-field over different latitude regions of the Sun and solar corona: i) equatorial or toroidal 
field (\emph{i.e.} the fields ranging in the latitude regions from $0^\circ$\,--\,$45^{\circ}$); ii) Mid-latitude fields (\emph{i.e.} the fields ranging in the latitude regions from $46^\circ$ - $78^{\circ}$; iii) the fields ranging from $0^\circ$\,--\,$78^{\circ}$
(henceforth referred it as Region-A fields); and iv) the polar or polar cap fields
(\emph{i.e.} the fields ranging in the 
latitude region from $78^\circ$\,--\,$90^{\circ}$). 
The averaged magnetic-fields measured over these latitude regions are 
shown in Figures \ref{fig:extrap1} and \ref{fig:extrap2}.
% \begin{itemize}
%  \item Equitorial or toroidal fields - $0^\circ$ and $45^{\circ}$
%  \item Mid-latitude fields - $46^\circ$ and $78^{\circ}$
%  \item Region-A fields - $0^\circ$ and $78^{\circ}$
%  \item Polar or Poloidal fields - $78^\circ$ and $90^{\circ}$
%  \end{itemize}

\subsection{Interplanetary Scintillations}\label{sec:ips_123}
If there is an enhancement or depletion of density fluctuations in the solar-wind along the 
line-of-sight (LOS) to the observed radio source, then there is a corresponding 
change in scintillation index [m] which defined as

\begin{equation}
 m={{\Delta S}\over{\langle S \rangle}},
\end{equation}

\noindent where $\Delta S$ and $\langle S \rangle$ are the scintillation flux and the 
mean source flux respectively. The quantity $\Delta S$ is computed from the power spectrum [$P(f)$] of the 
IPS observation using 
\begin{equation}
\Delta S =  \int_{0}^{\infty} P(f) df.
\end{equation}

\noindent The mean source flux ($\langle S \rangle$) can be measured by averaging the difference between onsource and offsource fluxes \citep{Tok2010}. IPS observations of 215 compact extragalactic and radio sources have been carried out on regular basis at 327 MHz. We would note here that all these sources have a finite angular diameter ranging from $\approx 10$ to $500$ milliarcsecond (mas) and observed over different heliocentric distances range from 0.2 to 0.8 AU. It was found that, in general, the quantity m increases with the decreasing heliocentric distance up to a certain distance called turn-over distance and beyond this distance it decreases rapidly with the further decrement in the distance. On the other hand, m decreases if the angular size of the radio source increases. Also, for an ideal point source the scintillating flux will be equal to the mean source flux and thus the scintillation index will be equal to unity at a certain distance (\emph{e.g.} at 327 MHz, the distance at which $m=1$ corresponds to 0.2 AU) and then decreases with the further increment in the heliocentric distance. Therefore, following \citet{Jan2011} and \citet{Bis2014}, we have normalized the m so that it is independent of the heliocentric distance and a finite source size at that distance. We would make a note that \citet{Jan2011} have reported the m after elimination of dependence of the  heliocentric distance but not corrected for the finite source sizes. In \citet{Bis2014}, the reported m is corrected for both dependence of heliocentric distance and the finite source sizes. 

For the sake of completeness we have summarized the method of normalization adopted in this article - (1) In order to remove the heliocentric distance dependence of m, each observation of m has to be normalized by that of a point source at that distance. We know that the source 1148-001 has the smaller angular diameter of $\lesssim 10$ mas at 327 MHz and therefore we treat that source to be an ideal point source \citep{Ven1985}. (2) Similarly, we have eliminated the dependence of finite source sizes using a least square minimization to determine which of the Marians curves best fits the data for a given source \citep[see][]{Mar1975, Bis2014}. Assuming the radio source 1148-001 as a point source, the observed values of m of all other sources were multiplied by a factor equal to the difference between the best fit Marians curve for the given source and the best fit Marians curve for 1148-001 at the corresponding heliocentric distance \citep{Bis2014}. Therefore, all measurements of m reported in this article were independent of a distance and the sources sizes. 

\begin{figure}[!ht]
   \centerline{\includegraphics[width=1.0\textwidth,clip=]{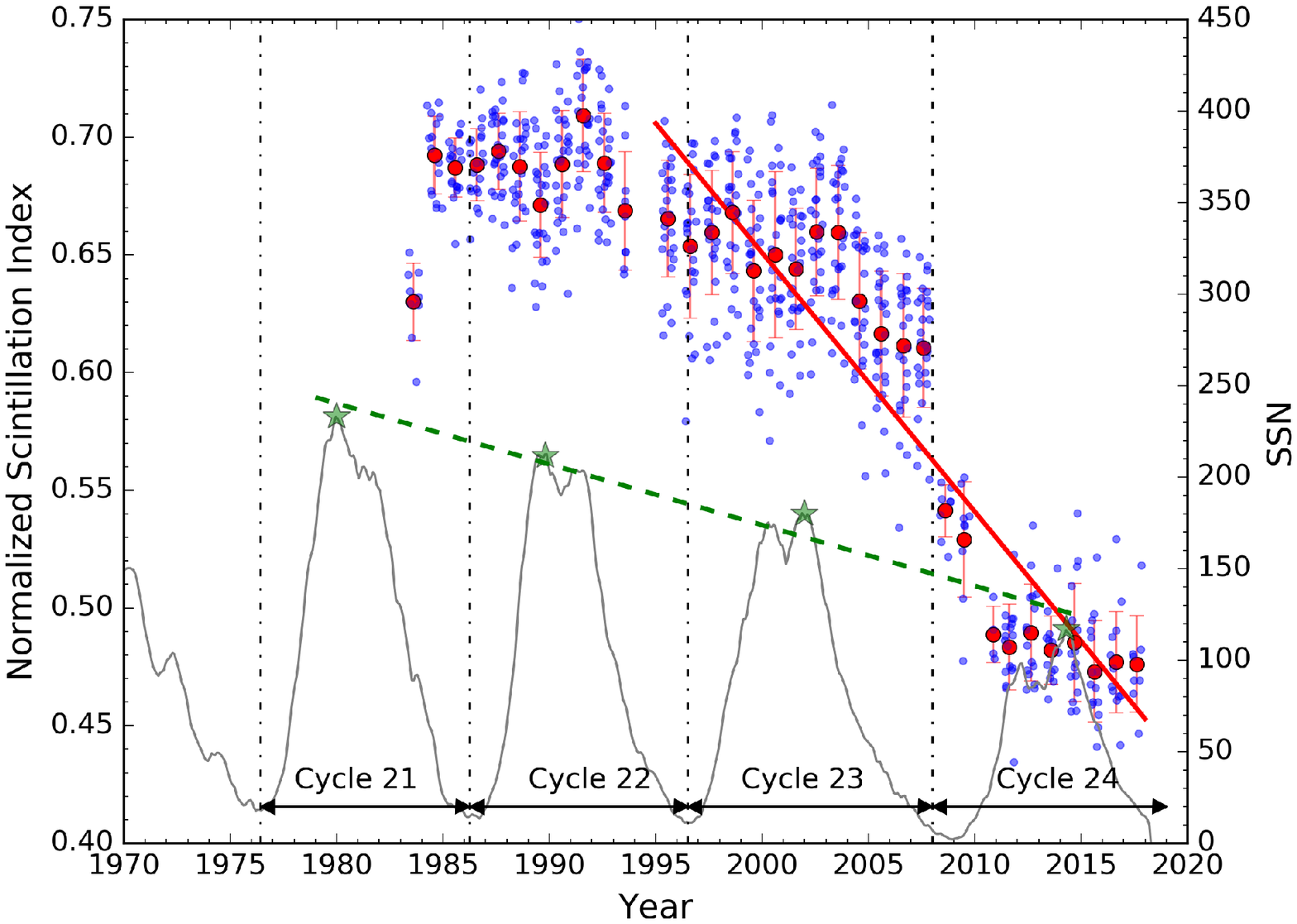}
              }
              \caption{Variation of the normalized scintillation index [$m$] over 
              different years is shown. The blue circles indicate the 
              annually averaged m for different sources. The \emph{red circles}
              indicate the annual average of the all sources observed in that year. 
              The \emph{fit to the red circles} (red solid curve) explicitly shows that 
              the quantity m has been declining since $\approx 1995$. These observations are 
              carried out using the facility at ISEE, Japan at 327 MHz during 1983\,--\,2017.
              The \emph{gray curve} indicate the sunspot number observed during this period. 
              The \emph{dashed green line} is a fit to observed highest sunspot number during the 
              solar maximum (indicated using \emph{green stars}) 
              also shows the declining highest sunspot number 
              since Cycle 21.\\
              }
\label{fig:m}
\end{figure}

After normalization, we have selected the sources which has at least 400 observations (during 1983-2017) and are uniformly distributed over the entire heliocentric distance (i.e., 0.2-0.8 AU) without a significant data gaps. After such a rigorous filtering, we have left with 27 sources (out of the 215 regularly observed sources) which cover the right ascensions over 24 hours and a wide range of declinations. The blue circles in the Figure \ref{fig:m} indicate the measured annual average of `m' that corresponds to 27 sources separately. The red circles indicate the annual average of m for the all 27 sources in a given year. We would like to make note that in order to avoid unusual error bars (during Solar Cycle 24) because of the significant drop in the m after 2008, we have measured the error bars separately for the years 1983-2008 and 2009-2017. The $1\sigma$ error bars of the annually averaged m values
of the observed sources in that year are shown in Figure \ref{fig:m}.

%We would make a note here that in the Figure \ref{fig:m}, we have eliminated the outliers whose standard %deviation is $> 2 \sigma$. 

\section{Results and Discussion}\label{sec:results}
Recent reports using angular-broadening observations show that 
various turbulent parameters (\emph{e.g.} amplitude of turbulence, 
density modulation index) correlate well with the solar 
cycle \citep{Bis2014, Sas2016, Sas2017}. In this article, we attempted to explore 
the relationship between the global magnetic-field strength and the 
turbulence parameters. As previously mentioned, there is no direct method
to measure the magnetic-field strength in the corona and hence in order to
have some idea of the magnetic-field strength at $10~R_{\odot}$ (the distance 
where, at present, we can probe the solar-wind using IPS observations), we used the 
PFSS extrapolation technique. We opt for this technique because it is basic 
and widely used in recent times. This technique is well approximated to the 
heliocentric distance up 2.5 $\rm R_{\odot}$. However, as the global magnetic-field 
lines are radial beyond this height, we extrapolated further to 10 $\rm R_{\odot}$.
As discussed in Section \ref{sec:data_analysis}, we inspected the magnetic-field
at photosphere, 2.5, and 10 $\rm R_{\odot}$ over different latitudes.

Figures \ref{fig:extrap1} and \ref{fig:extrap2} show the variation of the global magnetic-field with the solar cycle. In Figure \ref{fig:extrap1} panels a, c, e and panels b, d, f show the average of Region-A (latitudes $0^{\circ}-78^{\circ}$) and polar or poloidal fields. Similarly, in Figure \ref{fig:extrap2}, panels a, c, e and panels b, d, f show the average of the field in equatorial or toroidal and mid-latitude fields respectively. In both 
Figures \ref{fig:extrap1} and \ref{fig:extrap2}, in each panel the triangles 
pointing upward and triangles pointing downward indicate the northern and  southern hemispheric fields.  The circles in black 
indicate the mean field of both northern and southern hemispheric fields
(\emph{i.e.} average of the points shown in triangles both upward and downward). 
In each panel, for reference, the smoothed, monthly average, 
sunspot number (SSN) is shown as a gray solid line (\url{http://www.sidc.be/silso/DATA/SN_ms_tot_V2.0.txt}). In both Figures \ref{fig:extrap1} and \ref{fig:extrap2}, 
panels a and b show the magnetic-field at photospheric heights.
Panels c and d show the extrapolated field at $2.5~\rm R_{\odot}$ and 
panels e and f show the extrapolated field at 10 $\rm R_{\odot}$.

From Figures \ref{fig:extrap1} and \ref{fig:extrap2}, it is clear that 
magnetic-field has been declining since mid 1990s. Various observationally
derived parameters are tabulated in Table \ref{tab:table-1}.
The latitude range over which the magnetic-field is averaged is shown in 
column 2. The year from when the beginning of significant decline of the field
is shown in column 3 and the corresponding field in column 4. Similarly
for the year 2018 the measured magnetic-field is shown in column 6. 
We measured the decrement in magnetic-field (in percent) from mid-1990s to 2018 
(at photosphere, 2.5, and 10 $\rm R_{\odot}$) and tabulated it in column 7. 
We found that magnetic-field over different range of latitudes and heliocentric distances
are declined by $11-22.2\,\%$. These results are very significant support for the conclusion 
that not only were the polar fields declining since mid-1990, as previously reported \citep{Jan2011, Jan2015} but the overall (global) coronal magnetic-field is also declining. 
We found that, in phase with the global magnetic-fields, the quantity m also declined by 
$23.6\,\%$ from the mid 1990s to 2017. These results show that the 
global magnetic-field is controlling the 
turbulence characteristics in the solar corona and solar-wind. 
%Also, we found that the derived magnetic-fields at three heights (photosphere, 
%2.5 and 10 $\rm R_{\odot}$) are inversely proportional to $r^3$. 
%This shows the reliability of the extrapolated magnetic-fields at the corresponding
%heights. 

Another notable observation is that we see an oscillation of magnetic-field 
at the photospheric height in correlation with the sunspot number 
(see panels a and b of Figures \ref{fig:extrap1} and \ref{fig:extrap2}).
We examined such variation (from solar maximum to minimum) of the mean magnetic-field 
(see Figures \ref{fig:extrap1} and \ref{fig:extrap2}) in each solar cycle 
(from Solar Cycle 22 to 24) and found that it varied by $\approx 5-10\,\%$ .
We found that such oscillating behavior (due to sunspots) disappears in the solar corona and solar-wind (see panels c, d, e and f of Figures \ref{fig:extrap1} and \ref{fig:extrap2}) and shows a clear monotonic 
decline of the coronal magnetic-field (by $11-22.2\,\%$).

We also would like to make a note that in Figure \ref{fig:m}, the quantity m (the red circles) shows a negative jumps near the solar maximum which can be interpreted as follows - \citet{Sas2016} have reported that the density modulation index (\emph{i.e.} $\epsilon_N=\Delta N/N$; where $\Delta N$ and N are the density fluctuations and the background density respectively) positively correlates with the solar-wind speed. For the sake of completeness we provide the explanation given by them as follows. It was reported that the $\epsilon_N$ positively correlates with the temperature of solar-wind protons \citep{Cel1987}. Also, it was found that at 1 AU, the proton temperature positively correlates with solar-wind speed \citep{Lop1986}. Taken together, authors have concluded that $\epsilon_N$
should be larger in the fast solar-wind and lower in the slow solar-wind. 
We also know that during the solar minimum the global magnetic-field is dipolar and therefore, 
during solar minimum higher latitudes drive the fast solar-wind (because of the dominant polar coronal 
holes) and drive the slow solar-wind near the low latitudes. On the other hand, during solar maximum, 
the global magnetic-field is multi-polar and thus drives the slow solar-wind in all helio-latitudes
as the polar coronal holes are not prevalent \citep{Mcc2000, Asa1998}.
Therefore, during solar maximum the slow solar-wind suggests the lower density modulation index which in-turn is proportional to the quantity m \citep[see][]{Bis2014} and hence the negative jumps during solar maximum is consistent with the earlier reports.

\begin{table*}[!ht]
\centering
\vspace{5px}
\begin{tabular}{ccccccc}
\cline{1-7}

& Latitude & \multicolumn{2}{c}{Epoch - I} & \multicolumn{2}{c}{Epoch - II} & Decrement \\

\cline{3-6}
No.& range & Year & Mean field [G] & Year & Mean field [G] & in [$\%$] \\ 

(1) & (2) & (3) & (4) & (5) & (6) & (7) \\
\cline{1-7}
\multicolumn{7}{c}{Photosphere}	\\		
\cline{1-7}
1 &	$0^\circ-78^\circ$  & 1992 & 9.98  &  2018 &  8.32   & 16.6 \\
2 &	$78^\circ-90^\circ$	& 1997 & 6.56  &	2018 &	5.48   & 16.5 \\
3 &	$0^\circ-45^\circ$	& 1992 & 13.28 &	2018 &	11.17	 & 15.9 \\ 
4 &	$45^\circ-78^\circ$	& 1992 & 5.48  &	2018 &	4.44	 & 19.1 \\
\cline{1-7}
\multicolumn{7}{c}{$2.5~\rm R_{\odot}$}	\\		
\cline{1-7}
1 &	$0^\circ-78^\circ$	& 1996 & 0.11 &	2018 &	0.09	 & 15.4 \\
2 &	$78^\circ-90^\circ$	& 1997 & 0.19 &	2018 &	0.15   & 20.6 \\
3 &	$0^\circ-45^\circ$	& 1994 & 0.08 &	2018 &	0.07   & 11.3 \\
4 &	$45^\circ-78^\circ$ & 1997 & 0.15 &	2018 &	0.12   & 19.8 \\
\cline{1-7}
\multicolumn{7}{c}{$10~\rm R_{\odot}$}	\\		
\cline{1-7}
	
1 &	$0^\circ-78^\circ$	& 1992 & $3.51\times 10^{-3}$ &   2018 &	$2.79\times 10^{-3}$   & 20.6 \\
2 &	$78^\circ-90^\circ$	& 1994 & $3.82\times 10^{-3}$ &	2018 &	$3.12\times 10^{-3}$	 & 18.2 \\
3 &	$0^\circ-45^\circ$	& 1992 & $3.39\times 10^{-3}$ &	2018 &	$2.64\times 10^{-3}$	 & 22.2 \\
4 &	$45^\circ-78^\circ$ & 1993 & $3.71\times 10^{-3}$ &	2018 &	$3.03\times 10^{-3}$	 & 18.4 \\
\cline{1-7}

\end{tabular}
\caption{The averaged magnetic-field strength at different epochs (mid 1990s and 2018)
over different latitudes are shown. The decrement of the magnetic-field (in $\%$) over these epochs are shown in column 7. The measurements are tabulated for photospheric height, $2.5$, and $10~\rm R_{\odot}$.}
\label{tab:table-1}
\end{table*}

\section{Summary and Conclusion}\label{sec:conclusions}

Using the synoptic magnetogram data observed using NSO/KP and NSO/SOLIS instruments 
during the period from 1975 to April 2018, we inspected the average magnetic-fields (at the photosphere) over different latitude ranges in both northern and southern hemispheres.
We have noticed that not only the polar magnetic-field, but the 
equatorial, mid-latitude, and fields in Region A are also declining, since the 
mid-1990s (see Figures \ref{fig:extrap1}, \ref{fig:extrap2}; and Table \ref{tab:table-1}).
Further, we have inspected the magnetograms extrapolated (using the PFSS method) 
to the heliocentric distances
to 2.5 and 10 $\rm R_{\odot}$ and they also show the same trend. 
We found that, during the period from the mid-1990s to April 2018, the magnetic-field 
over different latitudes (at photosphere and inner solar-wind) declined by $11.3-22.2\,\%$. 
%The magnetic-field at 2.5 and 10 $\rm R_{\odot}$ shows approximately the way
%it plays a role in the solar-wind, which is difficult to study as there is no
%technique available to probe the magnetic-field directly at these heights.
Using the data observed from ISEE, Japan, we inspected the normalized scintillation index [$m$] during 1983\,--\,2017. We found that the quantity m has decreased by $23.6\,\%$ since the mid-1990s. From Figure \ref{fig:m}, it is clear that the peak sunspot number from 
Solar Cycle 21 (in the year 1980) to Solar Cycle 24 (in year 2014) has declined by $\approx 50\,\%$.
Therefore, these results show a strong relationship between the global magnetic-fields
and the various turbulence properties in the solar-wind.
Also, we found that magnetic-field at relatively low heights shows a monotonic decrease (by $15.9-19.1\,\%$; see Table \ref{tab:table-1}) 
as well as a variation of the magnetic-field (due to sunspots) over each solar cycle by $5-10\,\%$.
Such oscillating behavior disappears in the inner solar-wind (i.e. at 2.5 and $10~\rm R_{\odot}$) and a clear monotonic 
decline of the magnetic-field is seen (by $11.3-22.2\,\%$). 
In this article we show that the global coronal magnetic-field of the Sun (and not just the polar fields) 
has monotonically decreased since (approximately) 1995. 
These results are significant, as many authors are 
predicting that we are tending towards the another 
``Maunder''-like minimum \citep{Jan2011, Jan2015, Jan2018, Pes2018}.
It would be interesting to further examine the relationship between the global (large-scale) magnetic-field and the properties of density turbulence (which are measured by IPS).
For example, the way magnetic-field and the turbulence properties in the solar-wind 
(\emph{i.e.} amplitude of the turbulence, density, velocity and 
magnetic-field fluctuations, dissipation scales, and heating rates \emph{etc.}) are
related \citep{Bis2014, Sas2016, Sas2017, Sas2019}. The recently launched 
Parker Solar Probe may provide valuable insights in understanding
the relationship between the magnetic-field and the turbulent parameters \citep{Fox2016}.

\acknowledgements
K.S. Raja acknowledges Marc L. De Rosa for his valuable suggestions related to the PFSS extrapolation technique. K.S. Raja acknowledges the financial support from Centre National d'\'{e}tudes Spatiales (CNES), France. This work utilizes SOLIS data obtained by the \emph{NSO Integrated Synoptic Program} (NISP), managed by the National Solar Observatory, which is operated by the Association of Universities for Research in Astronomy (AURA), Inc. under a cooperative agreement 
with the National Science Foundation. Data storage supported by the University of 
Colorado Boulder ``PetaLibrary.'' Sunspot data from the World Data Center SILSO, Royal Observatory of Belgium, Brussels. The authors would like to thank the anonymous referee for his/her constructive suggestions and comments that helped in improving the manuscript. 
\bibliographystyle{aasjournal}
\bibliography{main}

\end{document}